\documentclass[aps,pra,reprint,amsmath,amssymb,showpacs,twocolumn]{revtex4}

\usepackage{bbm}
\usepackage{graphicx}
\usepackage{amsmath}
\usepackage{amssymb,amsthm}
\usepackage{color}
\usepackage[sans]{dsfont}

\newcommand{\ket}[1]{|#1\rangle}

\newcommand{\bra}[1]{\langle#1|}

\usepackage{hyperref}

\def\eq{\begin{eqnarray}}
\def\en{\end{eqnarray}}

\usepackage[normalem]{ulem}

\begin{document}

\title{Non-adiabatic many-atom quantum state control in  few-well  systems}
\author{Malte C. Tichy, Mads Kock Pedersen, Klaus M\o{}lmer, Jacob F. Sherson}
\affiliation{AU Ideas Center for Community Driven Research, CODER, \\
 Department of Physics and Astronomy, University of Aarhus, DK--8000 Aarhus C, Denmark}

\date{\today}
\begin{abstract}
We present a fast scheme for arbitrary unitary control of interacting bosonic atoms in a double-well. 
Assuming fixed inter-well tunnelling rate and intra-well interaction strength, we control the many-atom state 
by a discrete sequence of shifts of the single-well energies. For strong interactions, resonant tunnelling
transitions implement beam-splitter U(2) rotations among  atom number eigenstates, which can be  combined and, thus, permit full controllability. 
By numerically optimizing such sequences of \emph{couplings at avoided level crossings} (CALC), we extend the realm of full controllability to a 
wide range of realistic interaction parameters, while we remain in the simple control space. 
We demonstrate the efficiency and the  high achievable fidelity of our proposal with non-adiabatic population transfer, $N00N$-state creation, a C-NOT gate, and a transistor-like, conditional evolution of several atoms.
\end{abstract}
\pacs{
05.30.Jp,    
03.75.Lm, 
05.60.Gg, 
37.10.Jk 
}
\maketitle

\section{Introduction}
The possibility to count and manipulate individual particles  makes ultracold atoms  in optical lattices a highly attractive system for quantum state engineering and quantum computation  \cite{Folling2007a,Weitenberg2011,Anderlini2007,Weitenberg2011a}. While the fundamental building blocks for the manipulation and detection of individual atoms have been established, the control over more complex quantum states in ultracold-atom systems remains a challenge \cite{DAlessandro2008}.

In the paradigmatic double-well system, proposals exist for 
certain tasks such as the full transfer of atoms between the wells \cite{Weiss2005,DeChiara2008}, and the creation of particular classes of entangled states \cite{Weiss2005,Weiss2007}, $N00N$-states \cite{Lapert2012} and squeezed states \cite{Julia-Diaz2012,juliadiaz2012B}. 
 \emph{Arbitrary state control} of the atomic occupation dynamics 
  is a desirable key component to realize quantum atomtronics   \cite{Ruschhaupt2004,Micheli2004,Stickney2007,Seaman2007,Pepino2009,Pepino2010,Benseny2010,Gajdacz2012},
but no universal protocol has so far been proposed for this endeavour. 

General proofs of controllability have been achieved from two different perspectives: Any $N$-dimensional quantum system can always be controlled by the decomposition of the desired unitary U($N$) operation in elementary U(2) blocks \cite{Reck1994,Tilma2002,Schirmer2002}.  On the other hand, 
 every kinematically allowed 
  transformation can be realized 
 via ``bang-bang'' control \cite{Viola1998,Burd2002,Brion2011},  i.e.~by alternating the time-evolution induced by  two sufficiently non-commuting Hamiltonians \cite{Schirmer2008}. The 
  \emph{existence} of a control scheme does not, however, provide a solution to the 
   practical task of 
   reliably and quickly finding 
   simple and robust control sequences. As a result, in a many-body context, optimal control often needs  to be combined with numerical approximation techniques such as tDMRG  \cite{Doria2011} or density-functional theory \cite{Castro2012}.

Here, we achieve perfect unitary control by combining the two above   approaches. We use  a numerically optimized sequence of \emph{couplings at avoided level crossings} (CALC) to  bridge the gap between abstract existence proofs for control sequences \cite{Schirmer2008,Brion2011} and their actual, reliable and robust implementation in many-particle few-well systems  \cite{Lapert2012,Weiss2007,Ziegler2011,Julia-Diaz2012}. In particular, we focus on the truly high-fidelity ($>0.999$) preparation of quantum many-body states as possible resources for quantum information technologies. As examples, we apply our approach to the non-adiabatic transport of $N$ atoms from one well to another, to the creation of $N00N$-states, and to the controlled dynamics of a  single-species atom-transistor. Randomly sampling over target states reveals that truly every quantum state can be created with realistic variations of our control parameters. 

\begin{figure}[h]
\center
\includegraphics[width=\linewidth,angle=0]{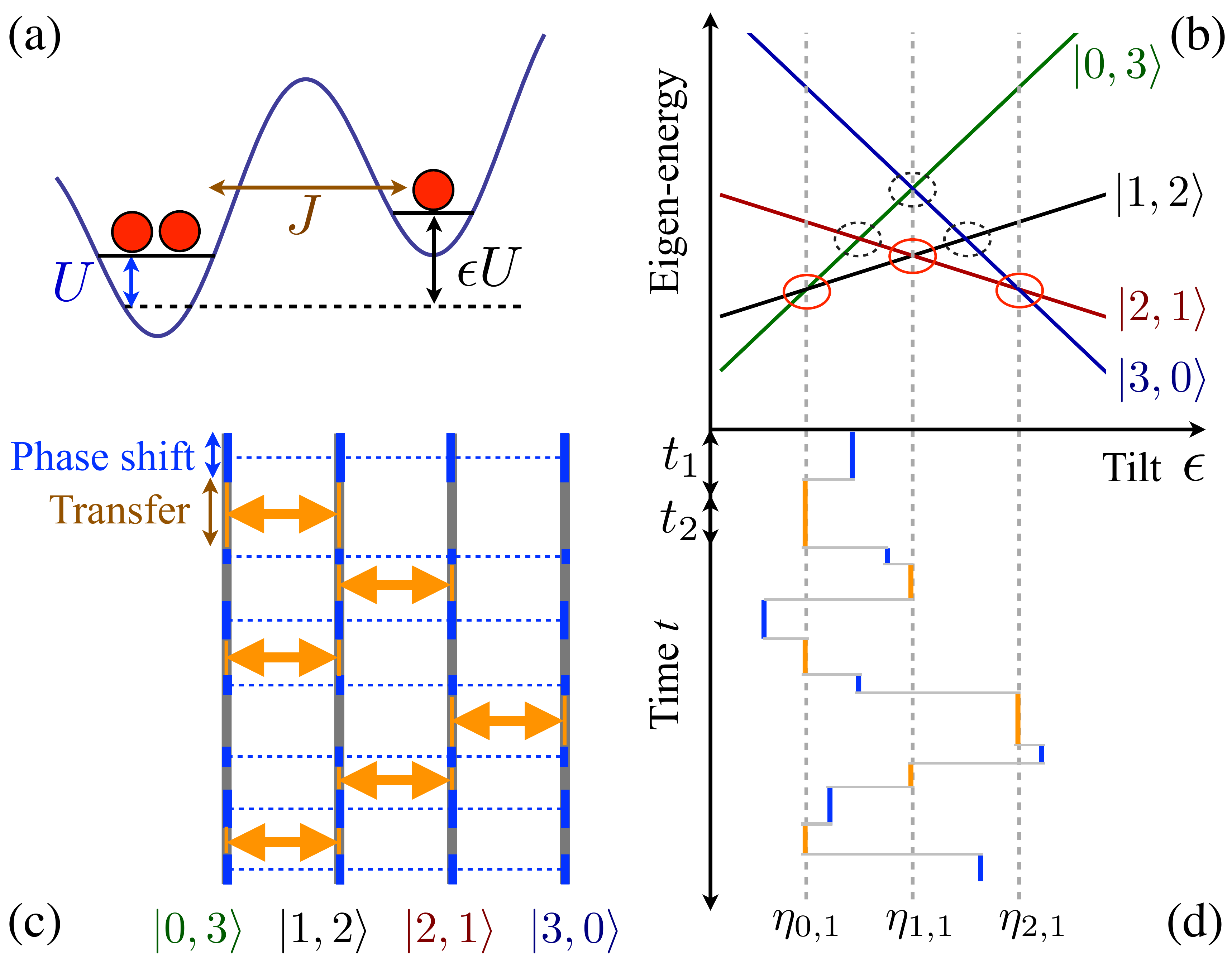}
\caption{(color online) (a) Bosonic atoms in a double-well system with adjustable tilt $\epsilon$.  (b) Fock-state energy levels as a function of the tilt  $\epsilon$ in the limit $U\gg J$, with highlighted first-order (red solid) and higher-order (black dotted) avoided level crossings. (c) Two-state couplers at avoided level crossings (large orange arrows) and phase-shifters at well-separated energies (blue dotted lines). (d) Time-dependent tilt $\epsilon(t)$ that realizes the couplings and phase shifts in the $4$-dimensional space of $N=3$ particles, as shown in (c). 
}
 \label{levelcrossingsbs}
\end{figure}

\section{Model and control strategy}

\subsection{Two-mode double-well model}
We consider a double-well system with $N$ bosons,  illustrated in Fig.~\ref{levelcrossingsbs}(a), which is described by the Bose-Hubbard Hamiltonian,
\eq
\hat H_{\text{BH}}(\epsilon ) &= & - \frac J 2 \left( \hat a^\dagger_1 \hat a_{2} + \hat a^\dagger_{2} \hat a_{1} \right) \label{BHH}  \\
\nonumber && + \frac U 2 \sum_{j=1}^2 \hat n_j (\hat n_j-1) +  \epsilon  \frac{U}{2} \left( \hat n_2 -\hat n_1 \right)  ,
\en 
where $U$ is the collisional interaction strength, $J$ is the inter-well tunnelling strength, and  
$\epsilon$ is the double-well tilt, proportional to the difference between single-particle energies in the wells. We assume that $\epsilon$ can be varied in the experiment \cite{Cheinet2008a,Simon:2011fk}. In the case of $U\gg J$ the $N+1$ two-mode Fock-states \eq \ket{N,0}, \ket{N-1,1}, \dots , \ket{0,N} , \label{fockstates} \en
constitute eigenstates of $\hat H_{\text{BH}}(\epsilon)$ for most values of $\epsilon$ (see Fig.~\ref{levelcrossingsbs}(b)).

Expanding  Eq.~(\ref{BHH}) in the Fock-state basis, one readily finds that, for $U\neq 0$, Hamiltonians associated with two different values of $\epsilon$ fulfill the requirements of ``bang-bang''-control  
 \cite{Schirmer2008,Brion2011}, such that 
  alternating applications of the two Hamiltonians can provide any unitary operation. 
   Already for $N=3$, however, the restriction to only two values of  $\epsilon$ is unnatural and inefficient, and in the following we shall suggest a simple control sequence that exploits the possibility for $\epsilon$ to attain a wide range of values.

\subsection{Strong-interaction limit}
We start by giving a systematic way to obtain the CALC sequence for strong interactions, $U \gg J$, for which the energy level crossings are well separated, as depicted in Fig.~\ref{levelcrossingsbs}(b). This sequence will then be used as a starting point for numerical optimization for realistic values of $U/J$. 

Two states $\ket{n,N-n}$ and ${\ket{n+m,N-n-m}}$ are degenerate  when 
\eq
\epsilon=\eta_{n,m} :=  2 n -N +m \  \ (n=0,\dots, N-1). \label{epsiloncond} 
\en 
\emph{Correlated} tunnelling of $m>1$ particles 
can only occur off-resonantly and it 
 is negligible in the regime $U\gg J$. Instead, single-particle tunnelling ($m=1$) dominates the picture: Given the tilt $\epsilon=\eta_{n,1}$, the 
Hamiltonian [Eq.~(\ref{BHH})] couples 
a degenerate pair of states with the bosonically enhanced frequency
\eq
\omega_n = J \sqrt{n+1} \sqrt{N-n}  ,
\en
while possible accidental degeneracies with $m>1$ can be neglected and 
the Hamiltonian remains diagonal in all the other Fock-states. 
 This immediately suggests that all atoms can be transferred from the right to the left well when the tilt is brought quickly from large negative values to $\eta_{0,1}$, then $\eta_{1,1}$ etc., spending the time
\eq
t_n=\pi/\omega_n \label{transport1} ,
\en
at each value $\eta_{n,1}$ to ensure complete state transfer at the avoided level crossing. In the limit $N \rightarrow \infty$, we  recover the continuous transfer protocol derived in Ref.~\cite{Weiss2005}.

Staying in the regime $U \gg J$,  a given duration spent at a first-order avoided level crossing leads to the equivalent of a beam splitter operation with a certain reflectivity on a pair of optical modes. A far-reaching result in quantum optics is that a general U($N+1$) multi-port beam splitter can be implemented by an arrangement of $N(N+1)/2$ two-mode beam splitters and $(N+1)(N+2)/2$ phase shifters \cite{Reck1994}.  Here, we require a maximum of $N-n$ operations on each pair of states $\ket{n,N-n}$ and $\ket{n+1,N-n-1}$, following the explicit decomposition of Ref.~\cite{Reck1994}. Exploiting these results for our system, we find that a CALC sequence that is constructed by strictly following the beam splitter analogy can implement every unitary operation in the limit $U\gg J$. Assuming that phase shifts, acquired at a rate proportional to $\epsilon U$, take a vanishing fraction of the time, we can give a bound on the required total protocol duration $T$ 
via  
   \eq T \le  \frac \pi J \sum_{j=0}^{N-1}  \sqrt{ \frac{ N-j }{j+1}}   . \label{totaltime} \en
   
\subsection{Optimization for finite interaction}
The validity of the above beam splitter analogy heavily relies on the assumption ${U \gg J}$. This physical regime is, however, not desirable in the laboratory, since inelastic collisional losses become important for strong interaction $U$, whereas the use of very small tunnelling strengths $J$ prohibitively increases the overall time-scale [Eq.~(\ref{totaltime})] of the process. Additionally, since tilts are effectively of the order of the interaction energy $U$ (see Eqs.~(\ref{BHH}) and (\ref{epsiloncond})), the employed tilts will necessarily invalidate the two-mode description for typical physical implementations of the double-well system when $U$ is no longer negligible in comparison to the energy gap to the first excited state. As a result, transitions  to higher bands will jeopardize the validity of the two-mode approximation. 

In Section \ref{tweezers} below, we will show that for a double-well realized by an optical double-tweezer, the two-mode description remains an excellent model for moderate values of $U/J \sim 2 \dots 10$.  However, in this regime, the avoided level crossings in Fig.~\ref{levelcrossingsbs}(b) become  broad, such that two-mode Fock-states cease  to be eigenstates of the problem. Additionally, higher-order avoided level crossings also become relevant. Consequently, by setting $\epsilon=\eta_{n,1}$, we not only induce transfer between $\ket{n,N-n}$ and $\ket{n+1,N-n-1}$ but also between other pairs of states that differ by one or several particles.  The simple picture of CALC dynamics as a sequence of beam splitters and phase-shifters is therefore not valid anymore. 

For finite $U/J$, an adiabatic change of the tilt from large negative to positive values induces the transition $\ket{0,N} \rightarrow \ket{N,0}$, via the slowly changing lowest energy eigenstate of the system \cite{Cheinet2008a}. The total protocol duration, however, is then much larger than $1/J$. In the following, we shall numerically optimize CALC sequences to account for finite $U/J$ and obtain a high-fidelity scheme that is universal and practical, in the sense that any target state $\ket{\Psi_{\text{target}}}$ can be reached starting with a given initial state $\ket{\Psi_{\text{initial}}}$ in few, simple operations.

A general control sequence  consists of $M$ values of the tilt $\epsilon_j$ ${(1\le j \le M)}$, which are each applied for a duration $t_j$. 
The sequence implements the unitary evolution 
\eq
\mathcal{\hat U}[ \vec t , \vec \epsilon] &=& \prod_{k=1}^M e^{- i \hat H_{\text{BH}}(\epsilon_k) t_k  } , \label{altunitary}
\en
where the product is understood to respect the time-ordering between the pulses. The fidelity of the state preparation is quantified by the overlap of the prepared state with the desired target state,
\eq
\mathcal{F} =| \bra{\Psi_{\text{target}} }  \mathcal{\hat U}[\vec t, \vec \epsilon ] \ket{\Psi_{\text{initial}}}|^2 . \label{Fidelity}
\en
The optimal vectors of time and tilt variables $(\vec t, \vec \epsilon)$ will not exactly coincide with the ones identified for ${U/J \rightarrow \infty}$ using the decomposition given in \cite{Reck1994}, but we find it convenient to use the CALC sequence as an initial guess for numerical optimization in which the pulse lengths $t_j$ as well as the tilts $\epsilon_j$ are adjusted, while the total number of steps $M$ is kept constant. For finite $U/J$, perfect fidelity can be achieved in most cases via such numerically optimized CALC sequences. When no satisfactory solution is found, the number of steps $M$ can be increased to extend perfect fidelity to the regime of even smaller $U/J$.

\subsection{Realisation with optical tweezers} \label{tweezers}
An experimental implementation of a double-well system should provide a physical realization of the Bose-Hubbard Hamiltonian [Eq.~(\ref{BHH})] that fulfils the following: On the one hand, the tilt $\epsilon$ should be tuneable at will, to permit control. On the other hand, the change of $\epsilon$ should not jeopardise the two-mode approximation; in particular, no transitions to higher bands should occur upon a change of $\epsilon$. In other words, the wavefunction of a particle that is initially prepared in the space spanned by the two energetically lowest single-particle eigenstates should remain in that space. 

A realization of such a versatile double-well system is given by the potential induced by two  optical tweezers with a  Gaussian beam profile, which lead to a trapping potential of the form \cite{Weitenberg2011a}
\eq 
V(x) = A_1 e^{-\frac{(x+b)^2}{2 c^2}} + A_2 e^{-\frac{(x-b)^2}{2 c^2}} ,  \label{tweezerpot}
\en
where $A_1$ and $A_2$ are controlled by the intensities of the two tweezers, $2b$ is the distance between the potential minima, and $c$ is the width of the tweezer potentials. We keep the distance between the tweezers constant with $b=0.25\mu$m. The effective barrier between the emerging wells (i.e. the ratio $U/J$) can then be controlled by the waist radius $c$, while the difference in the local energy $\epsilon$ is accessible by tuning the relative strength of $A_1$ and $A_2$. 

\begin{figure}[t]
\center
\includegraphics[width=\linewidth,angle=0]{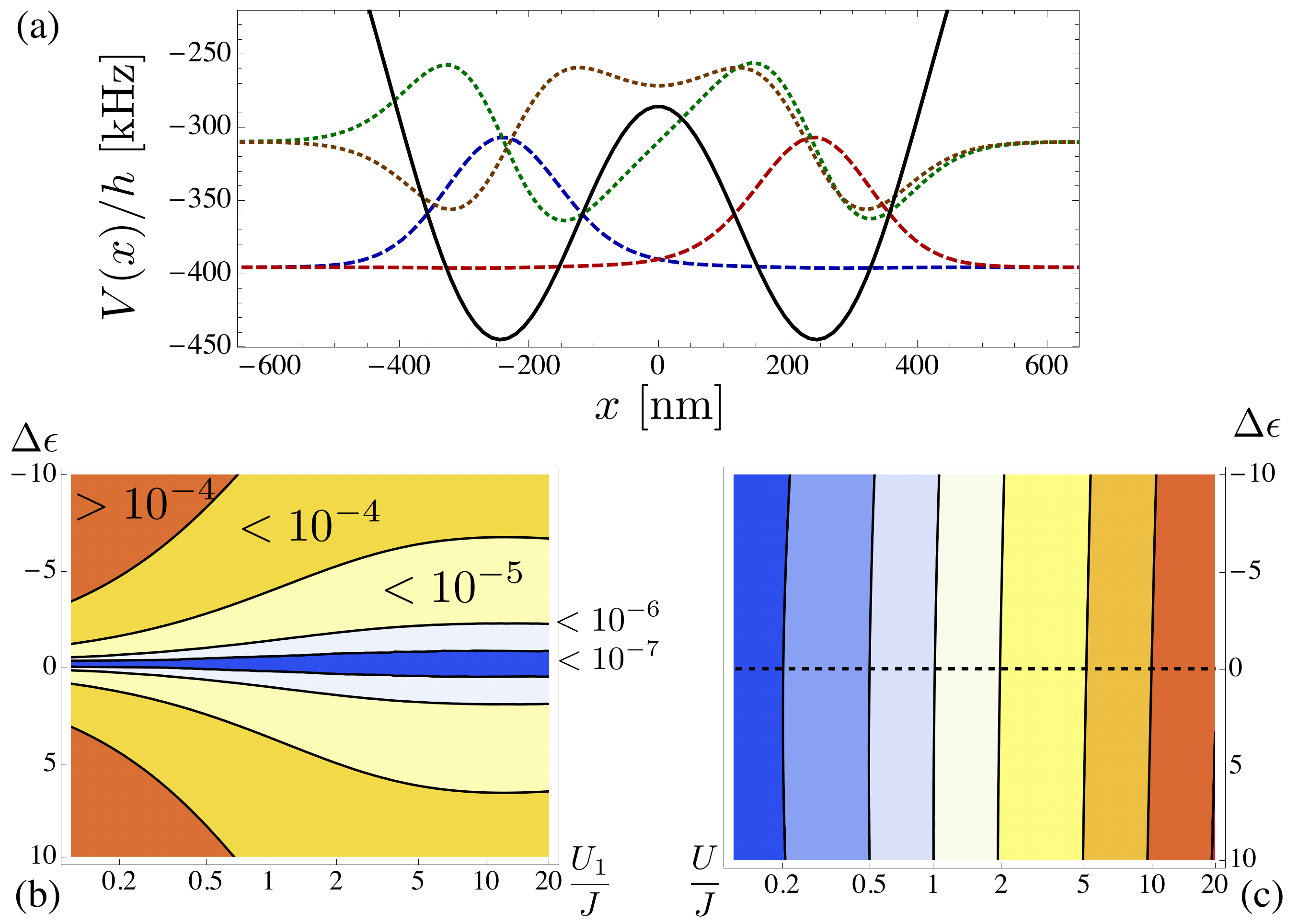}
\caption{
(color online) (a) Solid black: Potential of the optical double-tweezer [Eq.~(\ref{tweezerpot})], for $A_1/h=A_2/h=$440~kHz, $c=2b/3$. Dashed: Lowest localized eigenstates. Dotted: Two first excited states. (b) Infidelity of two-mode approximation [Eq.~(\ref{BHH})] under changes of the tilt from $\epsilon=0$ to $\epsilon=\Delta \epsilon$, as a function of $U_1/J$, where $U_1$ is the resulting interaction strength in the left well.  The solid lines separate areas that lead to the indicated infidelity. (c) Effective $U_1/J$ as a function of $U/J(\epsilon=0)$ and of the applied tilt $\Delta \epsilon$. A change of the tilt does not change the interaction strength significantly. The solid lines show paths with constant $U_1/J$. }
 \label{PotentialFig}
\end{figure}

The potential and its lowest eigenfunctions are shown in Fig.~\ref{PotentialFig}(a). The symmetric and anti-symmetric superpositions of the first two eigenfunctions of the full single-particle Hamiltonian yield the localized wave-functions in the left and the right well, respectively. Changes from $\epsilon=0$ to $\Delta \epsilon$ of the order $~N$ then lead to only slightly different eigenfunctions: The overlap of the two localized wavefunctions before and after a change of magnitude $\Delta \epsilon$ can be observed in Fig.~\ref{PotentialFig}(b). In our regime of interest, $U/J \sim 2 \dots 10$, the fidelity of the two-mode approximation remains higher than $1-10^{-4}$. In principle, changes of $\epsilon$ also induce weak changes in $U/J$, the resulting $U_1/J$ (the effective interaction for the left well eigenfunction) is shown in Fig.~\ref{PotentialFig}(c) for an $s$-wave scattering length of $100~a_0$. It remains widely independent of $\epsilon$.

\section{Applications}
\subsection{Non-adiabatic transfer}
\begin{figure}
\center
\includegraphics[width=\linewidth,angle=0]{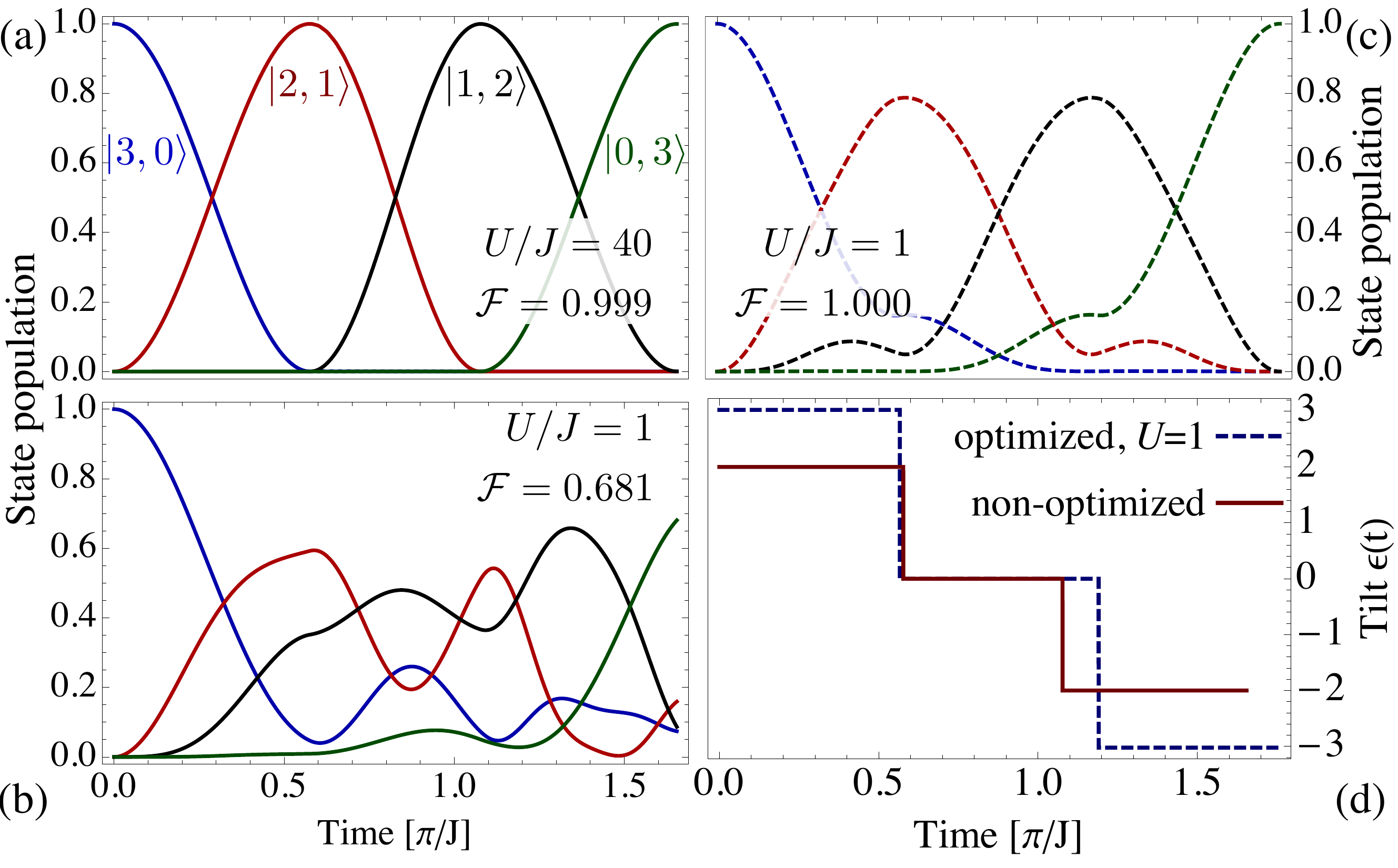}
\caption{(color online) (a) Target-state infidelity $1-\mathcal{F}$ for full population transfer $\ket{0,N} \rightarrow \ket{N,0}$ without optimization, as a function of the interaction strength $U/J$, for $N=2,3,5,10$ (red dotted, green dot-dashed, black solid, and blue dashed line, respectively). Optimizing numerically always yields arbitrarily high fidelity, better than $1-10^{-8}$, for any value of $U/J$ (not shown). (b) Total protocol duration for the numerically optimized population transfer. The thin dashed lines correspond to the ideal total protocol duration in the limit $U/J\rightarrow \infty$, $T=\sum_{k=1}^{N} t_k$, with $t_k$ given by Eq.~(\ref{transport1}). }
 \label{FidelitiesTranAndN00N}
\end{figure}

\begin{figure}
\center
\includegraphics[width=\linewidth,angle=0]{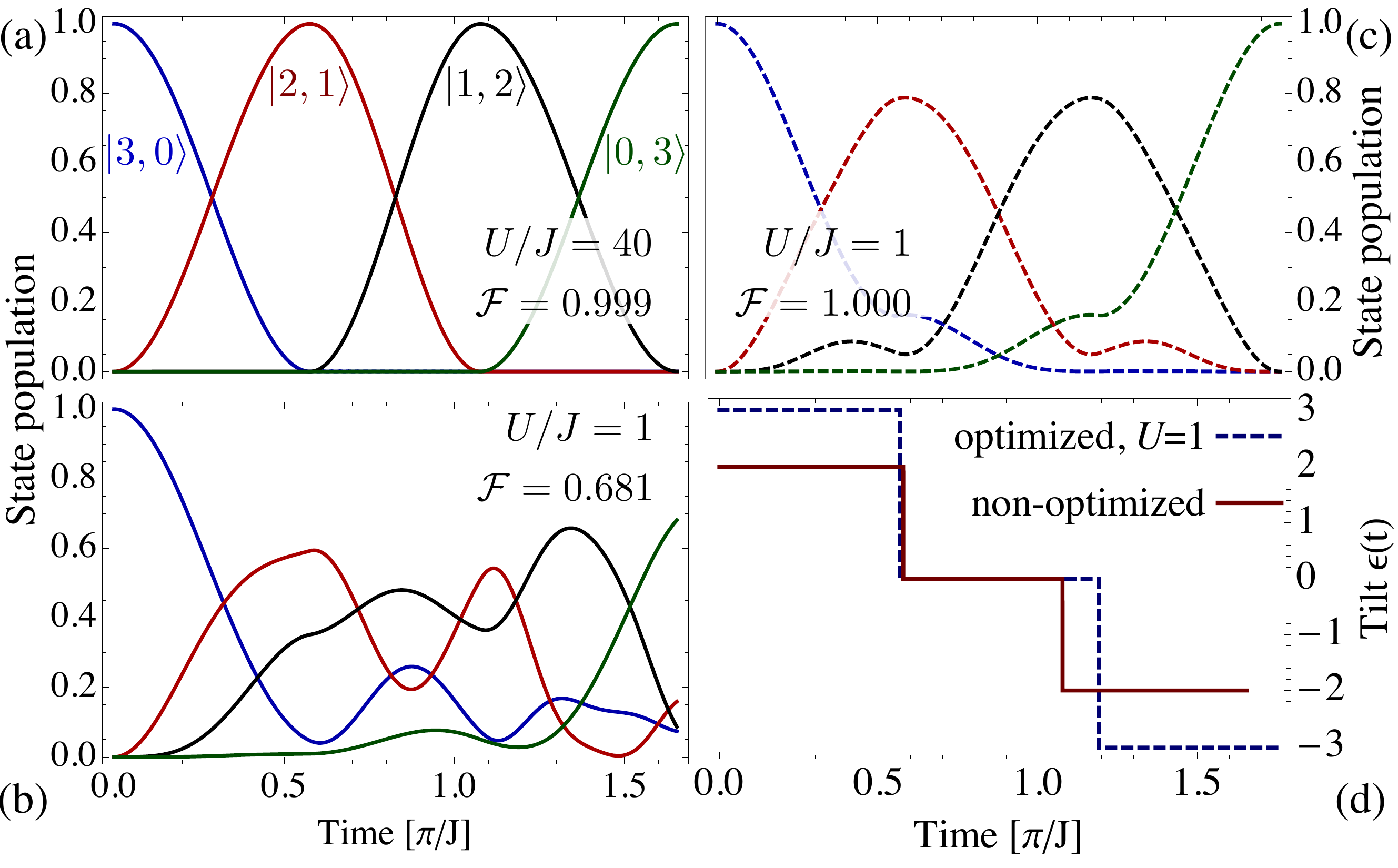}
\caption{(color online) Optimized and non-optimized sequences and population of state vectors for the full transfer of $N=3$ particles from the left to the right well, starting with $\ket{3,0}$. (a-c) Population of the different Fock-states and (d) pulse sequence, as a function of time. (a) The interaction $U/J=40$ is large, such that a non-optimized control sequence leads to a high fidelity. (b) The same control sequence, however, transfers the atoms only with around 68\% fidelity when the interaction is small, $U/J=1$. (c) By optimizing the control sequence numerically, perfect fidelity can be achieved. (d) The optimized pulse-sequence (dashed blue) differs from the one obtained analytically (solid red). The total time required for the optimized transfer is only slightly larger than for the non-optimized sequence.}
 \label{TransferAll.pdf}
\end{figure}

As a first application, we discuss the full transfer of $N$ particles between the  wells. Here, the unoptimized CALC sequence with $M=N$ resonant tilt values that are applied for durations given by Eq.~(\ref{transport1}) only leads to acceptable fidelities when the interaction $U/J$ is large and the number of particles small, as shown in Fig.~\ref{FidelitiesTranAndN00N}(a). 
Optimizing the times $t_k$, or, alternatively, the tilts $\epsilon_k$ does not lead to a satisfactory solution, but by simultaneously optimizing \emph{both} sets of parameters using a standard NelderÐ-Mead simplex algorithm \cite{NelderMead}, we quickly reach, for any value of $U/J$, an infidelity that is only constrained by numerical accuracy. We show the evolution of the wave-function components for $N=3$ in Fig.~\ref{TransferAll.pdf}. In (a), the interaction $U/J=40$ leads to a high fidelity, without optimization of the CALC sequence. Using the same sequence, however, an unacceptable, low fidelity is obtained for small interactions, $U/J=1$, in panel  (b). Optimizing, high fidelity is again recovered by effectively combining single-particle and correlated tunnelling through different intermediate states [panel (c)]. The unoptimized (solid red) and optimized (dashed blue) CALC sequences [panel (d)] are very similar, and, as shown in Fig.~\ref{FidelitiesTranAndN00N}(b), the total protocol durations of the unoptimized and of the optimized sequences do not deviate considerably. The fidelity of the process  depends on the accuracy of the time and tilt settings; numerical simulation reveals that $\mathcal{F} >0.999$ is reached as long as the times $t_k$ and the tilts $\epsilon_k$ are subjected to an error up to $1\%$ and $0.01$, respectively. 

 Although, for $U=0$, the trivial solution to the transport problem consists in setting $\epsilon_0=0$ for a time $t_0=\pi/J$ so that all particles tunnel at the same time, this protocol is not robust against small deviations from $U=0$, and even for small values of $U/J$, the  CALC sequence   provides a better starting point for optimization than the singular solution at $U=0$. 

While  unconstrained numerically optimized CALC sequences  are efficient for any $U/J$, the ``bang-bang'' 
  approach with only two different values of $\epsilon$ leads to considerably more complicated protocols: For instance, we needed 40 steps to achieve $\mathcal{F} \ge 0.99$ for the full transfer of $N=5$ particles at $U/J=5$, compared to the $M=5$ steps needed by CALC, and the number of steps increases even further for larger $U/J$.

\begin{figure}[ht]
\center
\includegraphics[width=\linewidth,angle=0]{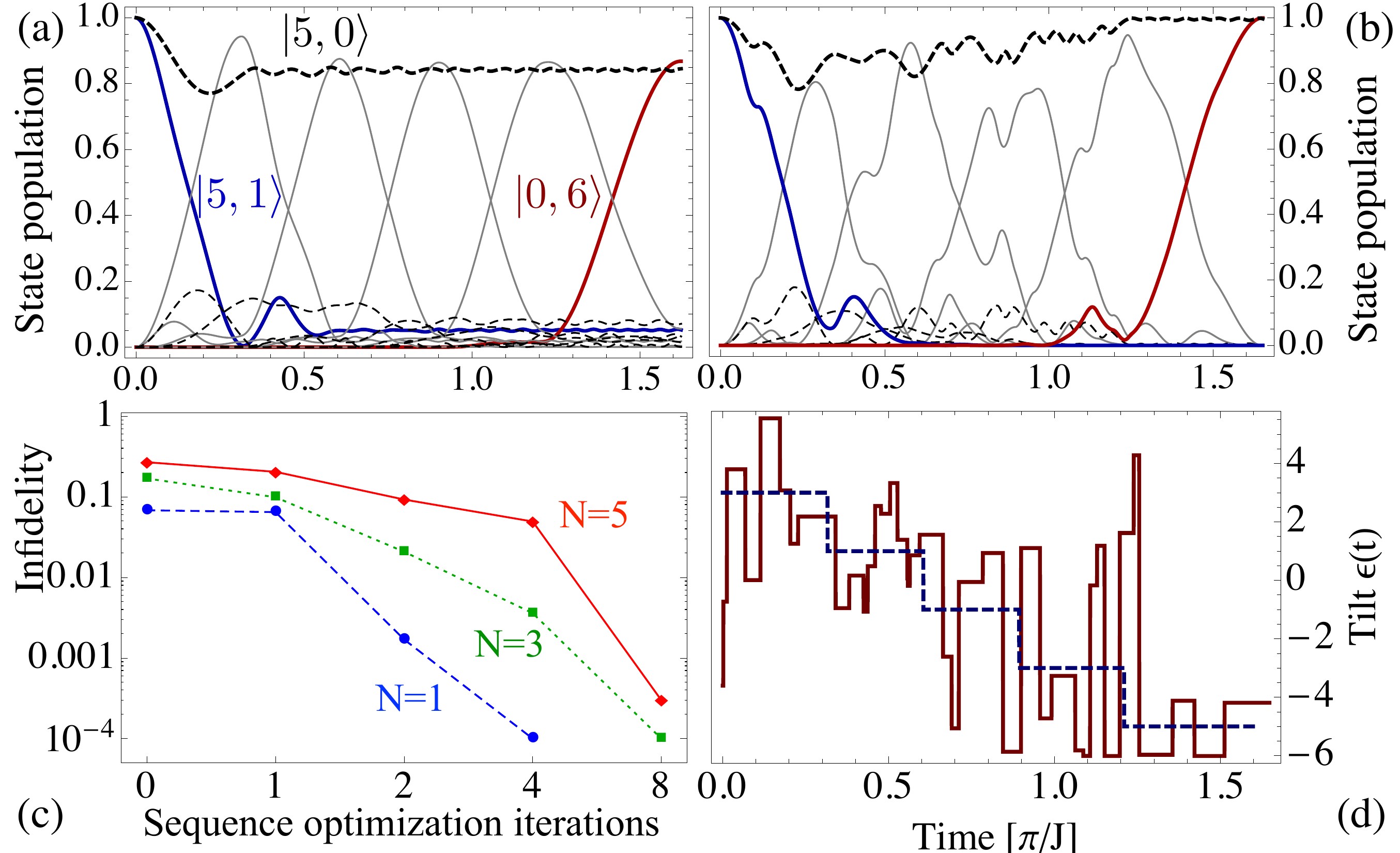}
\caption{(color online) Target-state infidelities $1-\mathcal{F}$ for $N00N$-state creation, $\ket{0,N} \rightarrow 1/\sqrt{2} \left(\ket{N,0} + e^{i \phi}  \ket{0,N} \right)$, without and with optimization, as a function of the interaction strength $U/J$. Blue dashed line: No optimization. Solid black line: Numerically optimized, for $M=N$. Red dotted line: Numerically optimized, for $M=2N$. To facilitate reading, we show the worst fidelity obtained for any interaction smaller than $U/J$ after optimization, which renders the curve monotonic. The achievable infidelity after optimization abruptly changes from an arbitrarily low value (here: $10^{-8}$, not shown) to a finite value at a certain $U/J$, depending on the number of particles. }
 \label{N00N23510}
\end{figure}

\subsection{N00N-state creation}
As an example of a non-trivial transformation, we consider the complete transfer protocol for $N$ atoms, but we maintain the first tilt $\epsilon=\eta_{0,1}$ for half the transfer time, $t_0=\pi/(2\omega_0)$, while the other tilts are kept as in Eq.~(\ref{transport1}). As can be readily understood from Fig.~\ref{levelcrossingsbs}(b), this sequence of $N$ steps transforms the Fock-state $\ket{0,N}$ into the entangled $N00N$-state, $(\ket{N,0}+\exp(i\phi)\ket{0,N})/\sqrt{2}$, where the relative phase $\phi$ can be adjusted by the time spent at tilt values away from the avoided level crossings. 
In Fig.~\ref{N00N23510}, we display the results for the generation of $N00N$-states. The dashed blue line represents the infidelity for the unoptimized  CALC sequence,  the solid black line shows the result after numerical optimization. 
 In contrast to the complete transfer of $N$ atoms between the wells, the infidelity of the optimized CALC sequence with $M=N$ steps rises abruptly below a critical value of $U/J$. This is consistent with the geometrical picture of SU(2) rotations around non-orthogonal axes \cite{DAlessandro2004,Chatzisavvas2009}: An operation that is composed of a finite fixed number of rotations around fixed Euler-axes in a high-dimensional space cannot produce every thinkable rotation when these axes are not orthogonal [an illustration is given in Figs.~2,3 of Ref.~\cite{Chatzisavvas2009}]. The larger the scalar product between the rotation axes, the larger will be  the necessary number of rotations around these axes that need to be concatenated to achieve every thinkable operation. Here, the rotation axes in the high-dimensional space are orthogonal in the limit $U/J \rightarrow \infty$, but they cease to be so for finite $U/J$. A mere doubling of the number of time-steps allowed in our optimization, $M \rightarrow 2M$, alleviates this loss of full controllability and significantly extends the range of $U/J$ that allows high fidelity $N00N$-state creation  (dotted red line). 

In general, we find that every two-mode state of $N$ atoms can be generated by numerically optimized CALC. This universality 
 was confirmed by picking a large random set of states 
that were all successfully generated  
 for finite $U/J$.

\begin{figure}[ht]
\center
\includegraphics[width=\linewidth,angle=0]{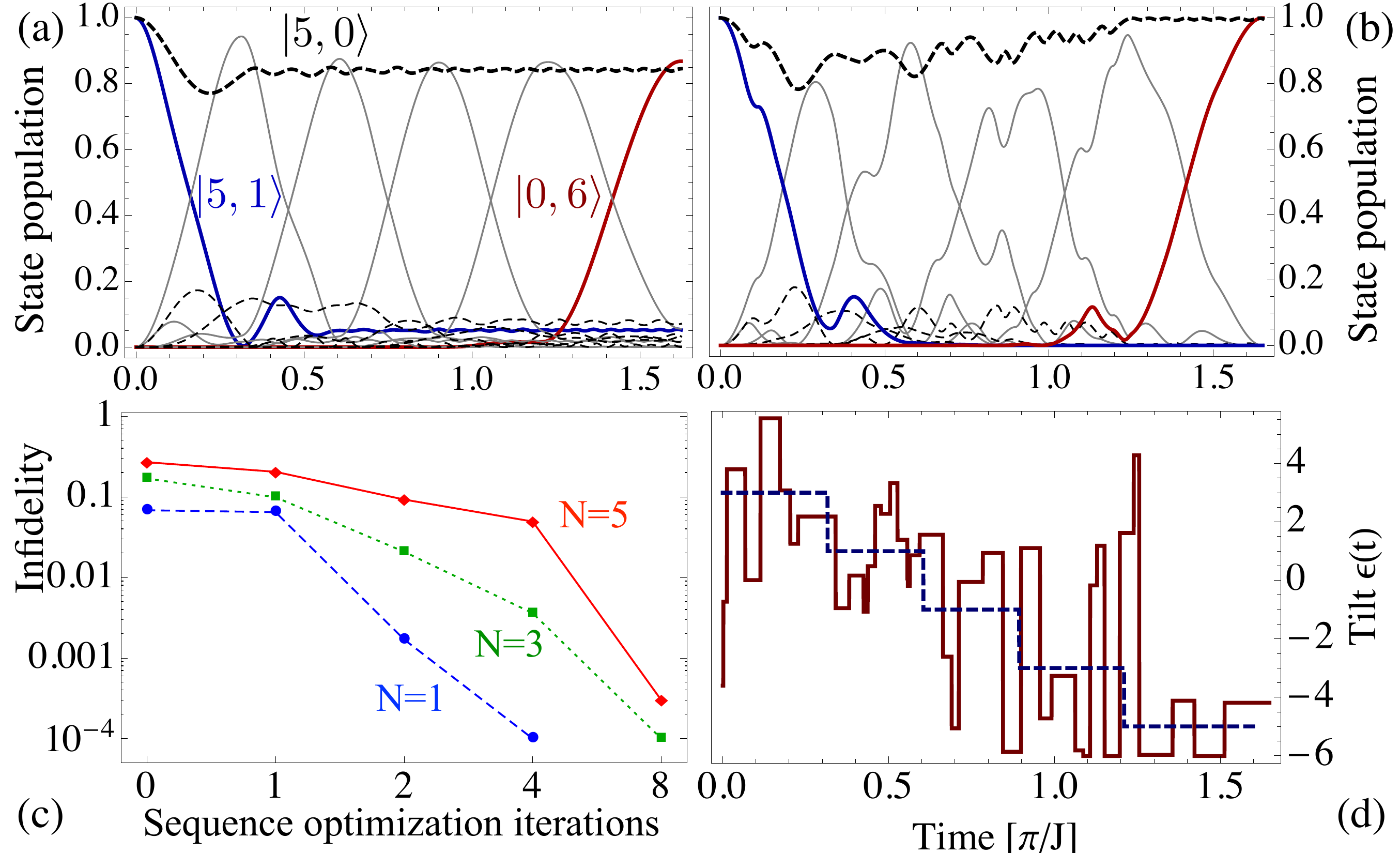}
\caption{(color online) Transistor-like time-evolutions. (a) Without optimization, the transistor sequence for $N=5, U/J=4$ only yields $\mathcal{F}\approx 0.73$: The $\ket{5,0}$-component (black dashed) is not eventually populated with unit probability, and a finite population stays in the $\ket{5,1}$-state (blue solid), at the expense of the $\ket{0,6}$-component (red solid). 
 Other components are drawn as thin gray ($N+1$-particle sequence) and black dashed lines ($N$-particle sequence).  (b) By optimizing the control sequence with 40 steps ($M=8 N$), $\mathcal{F} \ge 0.999$ is reached. (c) Infidelity of transistor-sequence for different numbers of particles and $U/J=4$, as a function of the number of iterations applied; 0 denotes no optimization, otherwise the number denotes the relative increase in the number of steps, $M/N$. (d) Non-optimized solution for $U \gg J$ (dashed blue) and optimized pulse-sequence for $U/J=4$ (solid dark red). }
 \label{transistor}
\end{figure}

\subsection{Single-atom transistor}
Finally, we apply numerically optimized CALC to a three-well single-atom transistor: Conditioned on the presence of an atom in the middle well of a three-well potential, all atoms in the left well are transferred to the right well. We assume that the single-particle energy in the right (left) well can be shifted with respect to the other two wells in the first (second) step of the procedure. The process can be implemented by two subsequent complete population transfer sequences that are tailored such that
\eq
\begin{array}{lclcl}
\ket{N,1,0} &\stackrel{(i)}{\rightarrow} & \ket{0,N+1,0} & \stackrel{(ii)}{\rightarrow} & \ket{0,1,N} \label{trans} , \\
\ket{N,0,0} & \rightarrow & \ket{N,0,0} & \rightarrow & \ket{N,0,0} \label{notrans}.
\end{array}
\en
The second step $(ii)$ is a special case of our double well control, and it only needs to be implemented successfully for the $N+1$ atoms present in the middle well in the first line, whereas the first step $(i)$ needs to be optimized to ensure that transfer only occurs into the already occupied middle well.
We assess the control sequence by the product of the fidelities of the two processes [Eq.~(\ref{trans})]. Due to the complexity of the dynamics, optimizing the CALC sequence while keeping the number of control steps $M=N$ does not yield acceptable fidelities. 
 By iteratively doubling the number of steps $M$ while keeping the total protocol duration approximately constant, we do, however, reach arbitrarily high fidelities for finite $U/J$, even under the additional constraint that the maximal tilt value $|\epsilon_j|$ remain always smaller than $N+1$ to exclude  higher-band heating effects. For $N=1$, the evolution [Eq.~(\ref{notrans})] can be used as the main building block of a two-qubit C-NOT-gate. We reach $\mathcal{F} \ge 0.999$ for all $U/J \ge 1$ with only $M \le 8$ settings of the tilt variables.  For $N=5$ and $U/J=4$, we show the unoptimized and the fully optimized time-evolution of the relevant components of the wave-function in Fig.~\ref{transistor}(a,b), and the evolution of the achievable fidelities with the number of iterations in Fig.~\ref{transistor}(c).  The minimal interaction strength required for $\mathcal{F} \ge 0.999$ at a constant number of steps $M$ increases with the number of particles, just like for the $N00N$-state creation above. Due to the complexity of the dynamics, errors in the time and tilt settings jeopardise the process fidelity more severely than for the two-well dynamics; numerical simulation of such errors showed that $\mathcal{F} > 0.99$ is obtained for $N=5$ when times (tilts) are accurate at the level of $1\%$  ($0.005$). We stress that unlike other atomic transistor proposals \cite{Gajdacz2012,Micheli2004} our approach does not involve two different atomic species.

\section{Conclusions}
Although any system with sufficient coupling between states permits full control \cite{Brion2011,Schirmer2008}, it is in many situations not clear how to  construct efficient and simple control sequences. In few-well systems with interacting bosons, the use of CALC sequences is a natural strategy that exploits the analogy to beam splitters in the limit of strong interactions. Numerically optimized CALC sequences can then achieve complex conditional dynamics, as required for quantum computation and atomtronics.  The CALC sequence for $U/J \gg 1$ can be easily obtained \cite{Reck1994}, it typically provides a fidelity above 90\% for desirable moderate values of $U/J\sim 2\dots 10$, which, on the one hand, is insufficient for applications, but, on the other hand,  provides a good starting point for numerical optimization to reach a fidelity that is arbitrarily close to unity. Our choice of optimization parameters is the simplest extension of the original  CALC sequence. Thus, thanks to the manageable, restricted number of optimization parameters, control via numerically optimized CALC is also feasible for complex Hamiltonians \cite{Zollner:2008uq}.  

A natural  extension is a protocol that tolerates particle number fluctuations: As long as the constraints of unitarity are not violated, there is no natural boundary for the design of complex conditional control sequences. Similarly, sequences that tolerate larger errors in the time and tilt sequences are desirable \cite{Montangero}. From a control-theory perspective, an extension of the SU(2) theory of generalized Euler-angles  for non-orthogonal rotation axes \cite{DAlessandro2004,Chatzisavvas2009} to larger spaces may complement our numerical results and give analytic insight in the minimal interaction strength $U/J$ that is required for perfect control under the constraint of a limited number of steps $M$. An interesting extension of the scheme described  by Eq.~(\ref{altunitary}) would be the adaptation to more sophisticated methods for ``shortcuts to adiabadicity'' \cite{Chen2010,juliadiaz2012B,13035615}.

\emph{Acknowledgements} The authors would like to thank Immanuel Bloch, Etienne Brion and Peter Schmelcher for helpful comments. M.C.T. gratefully acknowledges support by the Alexander von Humboldt--Foundation through a Feodor Lynen Fellowship.

\end{document}